\documentclass[ste,twoside]{stefano}

\usepackage{amsmath}
\usepackage{amssymb}
\usepackage{graphics}
\usepackage{rotating}

\def\makeheadbox{{%
\hbox to0pt{\vbox{\baselineskip=10dd\hrule\hbox
to\hsize{\vrule\kern3pt\vbox{\kern3pt
\hbox{  {\sf Modern Physics Letters} A {\bf 15}, 2057--2068 (2000) }
\hbox{  hep-ph/9906460 \hspace*{11.5cm}
$\boldsymbol{\Sigma \delta \Lambda}$ }
\kern3pt}\hfil\kern3pt\vrule}\hrule}%
\hss}}}

\def\mi{\mbox{\tiny $-$}}
\def\0{\mbox{\tiny $0$}}
\def\1{\mbox{\tiny $1$}}
\def\2{\mbox{\tiny $2$}}
\def\3{\mbox{\tiny $3$}}
\def\4{\mbox{\tiny $4$}}
\def\5{\mbox{\tiny $5$}}
\def\6{\mbox{\tiny $6$}}
\def\7{\mbox{\tiny $7$}}
\def\8{\mbox{\tiny $8$}}
\def\9{\mbox{\tiny $9$}}
\def\m{\mbox{\tiny av}}
\def\n{\mbox{\tiny $n$}}
\def\a{\mbox{\tiny $\alpha$}}
\def\b{\mbox{\tiny $\beta$}}
\def\at{\mbox{\tiny $at$}}
\def\so{\mbox{\tiny $so$}}
\begin{document}


\title{Remarks upon the mass oscillation formulas}

\author{
Stefano De Leo\inst{1}, 
Gisele Ducati\inst{2} \and
Pietro Rotelli\inst{3} 
}

\institute{
Department of Applied Mathematics, University of Campinas\\ 
PO Box 6065, SP 13083-970, Campinas, Brazil\\
{\em deleo@ime.unicamp.br}
\and
Department of Mathematics, University of Parana\\
PO Box 19081, PR 81531-970, Curitiba, Brazil\\
{\em ducati@mat.ufpr.br}
\and
Department of Physics and INFN, University of Lecce\\ 
PO Box 193, I 73100, Lecce, Italy\\
{\em rotelli@le.infn.it}
}


\date{Accepted {\em October 20, 2000}}

\abstract{
The standard formula for mass oscillations is often based upon 
the approximation $t \approx L$ and the  hypotheses that neutrinos 
have been produced with a definite momentum $p$ or, alternatively, 
with definite energy $E$. 
This represents an inconsistent scenario and gives an 
unjustified reduction by a factor of two in the mass oscillation 
formulas. 
Such an  ambiguity 
has been a matter of speculations and mistakes  in discussing 
flavour oscillations. We present a series of results and show
how the problem of the factor two in the oscillation length is not a 
consequence of gedanken experiments, i.e. oscillations in time.
The common velocity scenario yields the maximum simplicity.}



\offprints{~Stefano De Leo.\\
The authors were partially suppoerted by FAPESP grant 99/09008-5 (S.D.L), 
CAPES fellowship (G.D.) and INFN (P.R.).}

\maketitle

\section{Introduction}

\noindent One of the most popular fields of research in particle physics 
phenomenology
 of the last decades has been, and still is, 
that of neutrino 
oscillations~\cite{PON,NUS,BIPO,KAY,BIPE,GIU,RIC,KIE,GRI,DEL,STO}. 
Publications in this field 
have accompanied an ever increasing and stimulating series of 
experiments involving either  solar, atmospheric or 
laboratory 
neutrinos~\cite{APO,AMB,FUK,ATH,LSND,EITEL,BOE,BOE2,BOE3,BOE4,KAR,K2K}. 
The vast majority of the theoretical 
studies consider the possibility of 
massive neutrinos distinct from the flavour eigenstates created in the 
various production processes. 
Neutrinos are not the only example of such a 
phenomenon. The first examples of flavour oscillations observed were in 
the the kaon system~\cite{KS1,KS2}  
where the strong interaction is involved in 
the particle creation. We shall continue to refer in this work to neutrinos, 
but the considerations  are quite general.
What we shall call the {\em factor two problem\ } in the neutrino oscillation 
formulas has been already observed and discussed in previous 
papers~\cite{FAC2a,FAC2b,FAC2c,FAC2d},  
but the situation is still surprisingly confused and probably 
still subject to argument.
We would like to close the question with this paper, but more
realistically, we shall simply contribute to the general debate.

In this work we neglect the Heisenberg uncertainty principle~\cite{KK}. 
Physically this means that the wave packet of the created particle is large 
enough for its 
mass eigenstates to be assigned (as a good approximation) definite 
four-momentum. Indeed, the wave packet must be so large as to allow us to
legitemately assume a plane wave phase factor for each mass eigenstate.

Definite four-momentum implies definite velicities $v_{\n}$, which are in 
general not the same. Now in the Lorentz phases we must eventually insert 
values for $x$ and $t$. Different velocities automatically imply different 
values of $x_{\n}/t_{\n}$.

At creation the mass-eigenstate wave packets must coincide if a definite 
flavor is created {\em instantaneously}. However, this ia a frame dependent 
condition and in any serious analysis, one must allow for a ``formation time''
during creation. This fact alone allows the introduction of diferent $t_{\n}$
since at the laboratory the trigger will correspond (except for the equal 
velocity case) to non-equivalent points of the individual wave packets. This
also means different $x_{\n}$, whichever is the triggering point of 
the wave functions.

Our approach treats on an equal footing space-time and energy-momentum. 
Derivations of the oscillation formula which do {\em not} do this abound in 
the litterature. The most common are of two classes, one which imposes
in some way (see the next section) a common value for $x/t$ notwihstnding 
different velocities. The other class employs, a non-Lorentz invariant 
phase such that only time appears (for which a unique value is assumed) 
and then selects the Lorentz frame which yields the ``desired'' 
result.

We start the next section with one of the most authorative demonstrations of
the standard oscillation formula which is based upon the approximation 
$x\approx t = L$. We {\em correct} this derivation and show that it doubles 
the oscillation phase. In section 3 a number of short calculations are 
presented, each valid for different assumptions for the space-time intervals 
and the energy-momentum eigenvalues.

We draw our conclusions in section 4, where we emphasize that only situations 
of exactly equal ultra-relativistic velocities and for $\Delta M \ll M_{\n}$
will the standard formula be obtained. The important historical role and 
as a source of both high and low energy experiments played by the neutral 
kaon sector is briefly discussed.

\section{Neutrino mixing}

To  focus the contents of this paper we begin by claiming that standard 
oscillation formula, for mixing between mass  eigenstates, is often based upon 
the approximation $t \approx L$ and the assumptions of 
definite momentum or definite energy for the neutrinos created. 

To explain and understand the common mistakes in oscillation calculations,
let us briefly recall the standard approach. 
The most important aspect of neutrino 
oscillations can be understood by studying the explicit solution
for a system with only two types of neutrinos.
For this two flavour problem, the flavour
eigenstates $| \nu_{\a} \rangle$ and $| \nu_{\b} \rangle$ are represented
by a coherent linear superposition of the mass eigenstates 
$| \nu_{\1} \rangle$ and  $| \nu_{\2} \rangle$. 
\begin{equation}
\left[ \begin{array}{c}
|\nu_{\a} \rangle \\
|\nu_{\b} \rangle
\end{array} \right] =
\left[ \begin{array}{rr}
\cos \theta &  \sin \theta \\
- \sin \theta &  \cos \theta 
\end{array}
\right]
\,
\left[ \begin{array}{c}
|\nu_{\1} \rangle \\
|\nu_{\2} \rangle
\end{array} \right]~.
\end{equation}
The time evolution of $|\nu_{\a} \rangle$ is  determined by solving  the 
Schr\"odinger equation for the $| \nu_{\1,\2} \rangle$ 
component of $| \nu_{\a} \rangle$ in the rest  frame of that component   
\begin{equation}
\label{lif}
| \nu_{\n} (\tau_{\n}) \rangle = 
\exp \left[ -iM_{\n} \tau_{\n} \right] \, | \nu_{\n} \rangle
~~~~~~~\mbox{\tiny $n=1,2$}  
\end{equation}
where $M_{\n}$ is the mass of $| \nu_{\n} \rangle $ and 
$\tau_{\n}$ is the time in the mass eigenstate frame. 
In terms of the time $t$ and the position $L$  
the Lorentz-invariant phase factor in 
equation (\ref{lif}) must be rewritten as
\begin{equation}
\label{lif2}
\exp \left[ -iM_{\n} \tau_{\n} \right]  = \exp 
\left[-i (E_{\n} t - p_{\n} L) \right]~,
\end{equation}
where $E_{\n}$ and $p_{\n}$ are the energy and the momentum of the mass 
eigenstates in the laboratory, or any other, frame.  
In the standard approach the above equation is followed by 
this statement:
In practice, the neutrino is extremely 
relativistic, so the evaluation of the phase factor of
equation (\ref{lif2}) is calculated by making 
the approximation  $t \approx L$.
Consequently, equation (\ref{lif2}) becomes in this approximation
\begin{equation}
\label{lif22}
\exp \left[ -i (E_{\n}  - p_{\n}) L \right]~.
\end{equation}
For example, in 
the latest presentation contained in the Review of Particle 
Physics~\cite{DP}, the flavor-oscillation probability 
reads
\begin{equation}
\label{sf}
P ( \nu_{\a} , \nu_{\b} ) \approx 
\sin^2 2\theta ~ \sin^2 \left ( \frac{L}{4E} \, \Delta M^2 \right )  
  \approx       \sin^2  2 \theta ~  
          \sin^2 \left ( 1.27 \,  
\frac{L \, [\mbox{\footnotesize Km}]}{E \, [\mbox{\footnotesize Gev}]} \,  
\Delta M^2  \, [\mbox{\footnotesize eV$^2$}]\right )~, 
\end{equation}
where $\Delta M^2 \equiv M_{\1}^2 - M_{\2}^2$. Equation (\ref{sf}) 
is obtained by calculating the phase factor for each mass eigenstate 
traveling in the $x$ direction, $\exp \left[ -i(E_{\n} t-p_{\n} L)\right]$,  
with the 
approximation $t \approx L$ and the assumptions that $| \nu_{\a} \rangle$ 
has been produced
with a definite momentum $p$, 
\[
E_{\n} = \sqrt{p^2 + M_{\n}^2} \approx p + \frac{M^2_{\n}}{2p}~,
\]
or, alternatively, with  a definite energy $E$, 
\[
p_{\n} = \sqrt{E^2 - M_{\n}^2} \approx E - \frac{M^2_{\n}}{2E}~.
\]
The phase factor of equation (\ref{lif2}) then reads
\begin{equation}
\label{lif3}
\exp \left[ - i \frac{M^2_{\n}}{2p} L \right]~~~\mbox{or}~~~
\exp \left[ - i \frac{M^2_{\n}}{2E} L \right]~.
\end{equation}
``{\em Since highly relativistic neutrinos have $E\approx p$, the phase 
factors in equation (\ref{lif3}) are approximately equal. Thus, it 
doesn't matter whether $| \nu_{\a} \rangle$ is created with definite 
momentum or definite energy.}'' - Kayser~\cite{DP}. Now this result is 
incorret as we shall show below. First note that $t=L$ implies for consistency 
\[\frac{p_{\1}}{E_{\1}}=\frac{p_{\2}}{E_{\2}}=1~,\]
which, if simultaneously applied, eliminates the phase-factor completely. 
Null phase factors, as in all cases of equal phase factors
for each mass eigenstate, preclude any oscillation phenomena.
{\em Nor is such an approximation justified within a more realistic
wave-packet presentation}~\cite{KAY}.   
Returning to the simplified plane wave discussion, 
one should simply write
\begin{eqnarray*}
 \exp \left[ -i(Et-pL) \right] & = &  
\exp \left[ -i \left( \frac{E}{v}-p \right) L \right] =
\exp \left[ -i \, \frac{E^2-p^2}{p} \, L \right] = 
\exp \left[ -i \frac{M^2}{p} L \right]~,
\end{eqnarray*}
which differs from equation (\ref{lif3}) 
by a {\em factor of two}  in the argument. This 
simply doubles the coefficient of $\Delta M^2$ in the standard oscillation 
formulas.

The above result has already been noted by Lipkin~\cite{FAC2a},
 who however 
observes this ambiguity for the case of equal 3-momentum of the neutrino
mass eigenstates, the ``non-experiments'' as he calls them, 
but not for his chosen equal energy scenario. 
{\em We believe that only experiment can determine if in a given situation 
the neutrinos are produced with the same momentum or energy or neither}. 
For this reason we wish to present below
the differences in the various assumptions which
are particularly significant for non-relativistic velocities, admittedly not
very practical for the neutrino but very important for the kaon system. 
In any case we emphasize that the 
fore mentioned factor two appears not only in the scenario of common momentum 
but also for the equal energy assumption or ``real experiments'' as
Lipkin~\cite{FAC2d} calls them.

In realistic situations the flavour neutrino is created in a wave packet at 
time $t=0$ and thus over an extended region. We simplify our discussion by 
ignoring, where possible, this localization but we must note that it is 
essential to give an
approximate significance to $L$, the distance from source to measuring 
apparatus, or $t$, the time of travel. 

\section{Time or space oscillations?}

Assume that the state $ |\nu_{\a} \rangle $ is created at 
$t=0$ in $x \approx 0 $.
Introduce the Lorentz invariant plane wave factor and apply them to the 
mass eigenstates at a later time $t$ and for position $x$. 
Since the neutrino is created over an extended volume and the apparatus 
cannot be considered without dimension, the interference effects will involve 
in general amplitudes of states with different time and 
distance intervals. Different time intervals $t_{\1} \neq t_{\2} $ 
may seem an unnecessary and  unphysical abstraction. 
However, it is needed for self-consistency. Even if 
in a given frame the creation is considered instantaneous it will not 
generally 
appear so for another observer, given the extended dimension of the wave 
function. For this latter observer there will exist times when the probability 
of measuring the created particle is between 0 and 1. 
This implies the introduction, in general,
of a time dependence for the growth of a wave function at each 
$x$ in all frames. 
Furthermore, if we fix $L_{\1} = L_{\2} \equiv L $, and have different
velocities, $v_{\1} \neq v_{\2}$,
we must necessarily allow for $t_{\1} \neq t_{\2}$.
All this does not mean that the cases listed below are equally realistic.

Within
the approximation of an effective one dimensional treatment, 
we consider three broad classes:

\vspace*{.2cm}

\noindent
\begin{tabular}{lrrr}
$\bullet$ Common momentum~,
&~$p_{\1} = p_{\2} = p$~,
&~$E_{\1} \neq E_{\2}$~,
&$v_{\1} \neq v_{\2}$~; 
\\
$\bullet$ Common energy~,
&~$E_{\1} = E_{\2}= E$~,
&~$p_{\1} \neq p_{\2}$~,
&$v_{\1} \neq v_{\2} $~;
\\
$\bullet$ Common velocity~,
&~$E_{\1} \neq E_{\2}$~,
&~$p_{\1} \neq p_{\2}$~,
&$v_{\1} = v_{\2}$~.
\end{tabular}
\\

\noindent The above cases by no means exhaust all possibilities but they 
are sufficient to cover almost all the assumptions made in the literature 
and lead to the subtle differences of the resulting formulas for 
$P( \nu_{\a} , \nu_{\b} )$
which we are interested in.

The space-time evolution of $|\nu_{\a} \rangle$ and $| \nu_{\b} \rangle$ is  
determined by the space-time development of the mass eigenstates
$| \nu_{\1} \rangle$ and  $| \nu_{\2} \rangle$. In the laboratory frame,
we have 
\[  
|\nu_{\n} (L_{\n},t_{\n}) \rangle  =  
\exp \left[ - i (E_{\n}t_{\n} - p_{\n}L_{\n}) \right]  |\nu_{\n} \rangle
~~~~~~~\mbox{\tiny $n=1,2$}~.
\]
Consequently, the probability of observing a neutrino of a different flavor is
\begin{equation}
\label{e12}
P (\nu_{\a} , \nu_{\b} ) 
 =  \sin^{\2} 2\theta ~ \sin^{\2} \left[  
\mbox{$\frac{1}{2}$} \left(
E_{\2}t_{\2} - 
E_{\1}t_{\1} - p_{\2}L_{\2} + p_{\1}L_{\1} 
\right)
\right]~.
\end{equation}
We can eliminate the space or time dependence in the previous formula by
using the relations
\[
L_{\1} = \frac{p_{\1}}{E_{\1}}t_{\1}~~~~~\mbox{and}~~~~~
L_{\2} = \frac{p_{\2}}{E_{\2}}t_{\2}~.
\]
For time oscillations, we have 
\begin{equation}
P (\nu_{\a} , \nu_{\b}) 
 =  \sin^{\2} 2\theta~ \sin^{\2} \left[ \frac{M_{\2}^{\2}}{2E_{\2}}t_{\2} - 
\frac{M_{\1}^{\2}}{2E_{\1}}t_{\1} \right]~,
\end{equation} 
whereas, for space oscillations, we obtain 
\begin{equation}
P (\nu_{\a} , \nu_{\b} ) 
 =  \sin^{\2} 2\theta~ \sin^{\2} \left[ \frac{M_{\2}^{\2}}{2p_{\2}}L_{\2} - 
\frac{M_{\1}^{\2}}{2p_{\1}}L_{\1} \right]~.  
\end{equation}

\subsection{Common momentum scenario}

\noindent For {\em common momentum} neutrinos productions, 
we shall cosidere two different 
si\-tua\-tions, common arrival time and fixed laboratory distance, 

\vspace*{.2cm}

\noindent
\begin{tabular}{lr}
$\ast$ $t_{\1}=t_{\2}=t$~,~~~~~&
 $P (\nu_{\a} , \nu_{\b} ) 
 =  \sin^{\2} 2\theta~ \sin^{\2} \left[ \displaystyle{\frac{t}{2} \, 
    \left( \frac{M_{\2}^{\2}}{E_{\2}} - \frac{M_{\1}^{\2}}{E_{\1}} \right)} 
    \right]~;$\\
$\ast$ $L_{\1}=L_{\2}=L$~,~~~~~&
$P (\nu_{\a} , \nu_{\b} ) 
 =  \sin^{\2} 2\theta~ \sin^{\2} \left[ \displaystyle{\frac{L}{2p} \, 
     \left( M_{\2}^{\2} - M_{\1}^{\2} \right)} \right]~.$  
\end{tabular}
\\

\noindent 
As already mentioned, for Lipkin \cite{FAC2d}
time oscillations represent non experiments or ge\-dan\-ken experiments 
because they measure time  oscillations.  
For ``real'' experiments, in the 
scenario of common momentum neutrino production, we should use the 
formula
\begin{equation}
\label{sop}
P (\nu_{\a} , \nu_{\b} ) 
 =  \sin^{\2} 2\theta~ \sin^{\2} \left[ \frac{L}{2p} \, \Delta M^{\2} 
    \right]~. 
\end{equation}
In terms of the average neutrino energy,
\[
E_{\m} = \frac{E_{\1} + E_{\2}}{2} = p \, 
\left[ 1 + \frac{M_{\1}^2 + M_{\2}^2}{4 p^2} + {\cal O} \left( \frac{M^4}{p^4}
\right) \right]~,
\]
we can rewrite the previous equation, in the ultra--relativistic limit
as 
\begin{equation}
\label{sop2}
P (\nu_{\a} , \nu_{\b} ) 
 \approx  \sin^{\2} 2\theta~ \sin^{\2} \left[ \frac{L}{2E_{\m}} \, 
\Delta M^{\2} 
    \right]~. 
\end{equation}
Thus, we find a factor two difference between 
the oscillation coefficient in this formula and 
the standard mass oscillation formula of equation (\ref{sf}).

\subsection{Common energy scenario}

\noindent Let us now consider {\em common energy} neutrinos productions,

\vspace*{.2cm}

\noindent
\begin{tabular}{lr}
$\ast$ $t_{\1}=t_{\2}=t$~,~~~~~&
$P (\nu_{\a} , \nu_{\b} ) 
 =  \sin^{\2} 2\theta~ \sin^{\2} \left[ \displaystyle{\frac{t}{2E} \, 
     \left( M_{\2}^{\2} - M_{\1}^{\2} \right)} \right]~;$\\ 
$\ast$ $L_{\1}=L_{\2}=L$~,~~~~~&
$P (\nu_{\a} , \nu_{\b} ) 
 =  \sin^{\2} 2\theta~ \sin^{\2} \left[ \displaystyle{\frac{L}{2} \, 
    \left( \frac{M_{\2}^{\2}}{p_{\2}} - \frac{M_{\1}^{\2}}{p_{\1}} \right)} 
    \right]~.$
\end{tabular}  
\\

\noindent Space oscillations are described by
\begin{equation}
\label{soe}
P (\nu_{\a} , \nu_{\b} ) 
 =  \sin^{\2} 2\theta~ \sin^{\2} \left[ \frac{L}{2} \, 
       \Delta \left( \frac{M^2}{p} \right) \right]~,
\end{equation}
where
\[ 
\Delta \left( \frac{M^2}{p} \right) = 
\frac{M_{\2}^{\2}}{p_{\2}} - \frac{M_{\1}^{\2}}{p_{\1}}~.
\]
This can be written as
\[ 
\Delta \left( \frac{M^2}{p} \right) =  
\frac{\Delta M^{\2}}{E} \left[ 1 + \frac{M_{\1}^2 + M_{\2}^2}{2 E^2} + 
 {\cal O} \left( \frac{M^4}{E^4} \right) \right]~. 
\]
Consequently, in the ultra-relativistic limit, equation (\ref{soe}) becomes
\begin{equation}
\label{soe2}
P (\nu_{\a} , \nu_{\b} ) 
 \approx  \sin^{\2} 2\theta~ \sin^{\2} \left[ \frac{L}{2E} \, \Delta M^{\2} 
    \right]~. 
\end{equation}
{\em The factor two difference is thus also 
present in common energy scenarios}.
The formulas in equations (\ref{sop}) and (\ref{soe}) tend to the same result 
in the ultra-relativistic limit, equations (\ref{sop2}) and (\ref{soe2}), and 
are in disagreement with the standard formula, equation (\ref{sf}).
In theory, at least the differences between them may be experimentally
determined, especially for non-relativistic processes.

\subsection{Common velocity scenario}

The scenario of different momentum and energy neutrinos productions,
with {\em common velocity}, merits special attention, because only if
$v_{\1} = v_{\2}$ the probability $P ( \nu_{\a} , \nu_{\b} )$ is 
valid for all times. Otherwise, $P ( \nu_{\a} , \nu_{\b} )$
is valid only until the wave packet for the two mass eigenstates overlap 
substantially. 
This complication does not exists for $v_{\1} = v_{\2} $. Indeed, with this 
condition, the wave packets travel together for all observers and we may 
even employ a common $L$ and common $t$. 
Furthermore, there exists in this case a rest frame,
$v=0$, for our flavour eigenstate common to that of the mass
eigenstates. This situation is implicit in all calculations that 
use a common proper time $\tau$.
By assuming a common velocity scenario,
we must necessarily require
different momentum and energies for the neutrinos produced. 
Due to the common 
velocity, the time evolution for the mass eigenstates in the 
common rest frame is
\begin{equation}
\label{lifc}
| \nu_{\n} (\tau) \rangle = \exp \left[ -iM_{\n} \tau \right] \, 
| \nu_{\n} \rangle~,
\end{equation}
and only in this case is equation (\ref{lif2}) really justified
with its non indexed time and distance. 
In fact, for common velocities, the Lorentz-invariant phase factor 
can be rewritten in terms of the common time, $t$, and the 
common position, $L$, in the 
laboratory frame. We can eliminate the time dependence in the previous 
formula by using the relations
\[
L = \frac{p_{\1}}{E_{\1}} \, t = \frac{p_{\2}}{E_{\2}} \, t~.
\]
Space oscillations, are thus described by 
\begin{equation}
P (\nu_{\a} , \nu_{\b} ) 
 =  \sin^{\2} 2\theta~ \sin^{\2} 
\left[ \frac{L}{2} \, \left( \frac{M_{\2}^{\2}}{p_{\2}} - 
\frac{M_{\1}^{\2}}{p_{\1}}\right) \right] ~.  
\end{equation} 
This equation is formally equivalent to equation (\ref{soe2}) and thus, at 
first glance,  it seems to reproduce the factor two difference.
{\em This is a wrong conclusion!} Indeed, in the scenario of common 
velocity,
\[ p_{\1} = M_{\1} \gamma_v v~~~~~\mbox{and}~~~~~p_{\2} = M_{\2} \gamma_v v~,\]
and consequently
\[
\Delta \left( \frac{M^2}{p} \right) = 
\frac{ M_{\2} - M_{\1}}{\gamma_v v} = \frac{\Delta M^2}{2 p_{\m}}~.
\]
Space oscillations, in the common velocity scenario, are thus
described by 
\begin{equation}
\label{soc}
P (\nu_{\a} , \nu_{\b} ) 
 =  \sin^{\2} 2\theta~ \sin^{\2} \left[ \frac{L}{4 p_{\m}} \, 
       \Delta M^2   \right]~,
\end{equation}
and this recalls the standard result with the factor four in the
denominator.
However, it must be noticed that for this case 
\[ \frac{E_{\1}}{E_{\2}}=\frac{p_{\1}}{p_{\2}} = \frac{M_{\1}}{M_{\2}}~,\]
and this may be
very far from unity. Thus, the use of $p_{\m}$ in equation (\ref{soc}) is not
exactly the $E$ intended in the standard formula, which was identical,
or almost, for both neutrinos.

We conclude our discussion  by giving the mass oscillation 
formula in terms of the ``standard'' phase factor
\[
\frac{L}{4 E_{\m}} \, \Delta M^2~,
\]
and the parameter 
\[
\alpha =  
\left( 1 + \frac{\gamma_{\1} M_{\1}}{\gamma_{\2} M_{\2}} \right)  
\left( 
\frac{1}{v_{\2}} - \frac{1}{v_{\1}} \frac{\gamma_{\2} 
M_{\1}}{\gamma_{\1} M_{\2}}
\right) \left( 1 - \frac{M_{\1}^2}{M_{\2}^2} \right)^{-1}~.
\]
The new formula reads
\begin{equation}
\label{eq}
P (\nu_{\a} , \nu_{\b} )  =  \sin^{\2} 2\theta~
 \sin^{\2} \left[\alpha \, \frac{L}{4 E_{\m}} \, \Delta M^2 \right]~.
\end{equation}
For common velocity, momentum and energy, the parameter $\alpha$ becomes 
\[
\begin{array}{lclcl}
\alpha_v & \equiv  &  \alpha \, [v_{\1}=v_{\2}]
          &=&  1/v~,\\
\alpha_p & \equiv &  \alpha \, [p_{\1} = p_{\2}] 
         & =&  
(v_{\1} + v_{\2})/ v_{\1} v_{\2}~,\\
\alpha_E & \equiv & \alpha \, [E_{\1}=E_{\2}] 
         & =&  
2 \, \left( 1 + \frac{1}{v_{\1} v_{\2}} \right) / (v_{\1} + v_{\2})~. 
\end{array}
\]
Finally, in the ultra-relativistic limit, by killing the 
${\cal O} \left( \frac{M^4}{p^4} \, , \,  \frac{M^4}{E^4}  \right)$ 
terms, we obtain 
\[
\begin{array}{lcl}
\alpha_v & \approx & 
1 + \left( M_{\1}^2 / p_{\1} + M_{\2}^2 / p_{\2} \right) / 
4p_{\m}~,\\
\alpha_p & \approx & 
2 + \left( M_{\1}^2 + M_{\2}^2 \right) / 2 p^2~,\\
\alpha_E & \approx &
2 + \left( M_{\1}^2 + M_{\2}^2 \right) / E^2~. 
\end{array}
\]

\section{Conclusions}

The creation of a particle may differ 
from process to process, therefore only 
experiment can decide which, if any, of the above situations are 
involved~\cite{WIN}. However, we wish to point out that
{\em the assumptions of same momentum, $p$, or same energy, $E$, 
can only be valid in, at most, one reference frame}. 
It seems to us highly unlikely that this frame happens to
coincide with our laboratory frame. This means that if the common 
velocity scenario, 
{\em which is frame independent}, is not satisfied, we may
legitimately 
doubt that any of the popular hypothesis coincide with any given 
experimental situation.

The plane wave treatment of mass oscillations gives the result,
\[
P (\nu_{\a} , \nu_{\b} )  \approx  \sin^{\2} 2\theta~
 \sin^{\2} \left[ \frac{L}{4 E_{\m}} \, \Delta M^2 \right]~,
\]
{\em under certain conditions}, i.e. the different mass eigenstates have
a common ultra-relativistic velocity, $v_{\1}=v_{\2}=v$,
\[
 P (\nu_{\a} , \nu_{\b} )  =  \sin^{\2} 2\theta~
 \sin^{\2} \left[ \frac{L}{4 v E_{\m}} \, \Delta M^2 \right]
\approx
\sin^{\2} 2\theta~
 \sin^{\2} \left[ \frac{L}{4 E_{\m}} \, \Delta M^{2} \right]
 ~.
\]
However, it immediately raises a number of conceptual questions. Why should 
the different mass eigenstates have a common velocity? 
We have shown in this work that, for 
ultra-relativistic neutrinos,
the scenario of common momentum or common energy {\em doubles}
the oscillation amplitude, yielding the standard oscillation formula
\[
P (\nu_{\a} , \nu_{\b} )  \approx  \sin^{\2} 2\theta~
 \sin^{\2} \left[ \frac{L}{2 E_{\m}} \, \Delta M^2 \right]~.
\]
{\em The difference between the scenarios of common momentum and common energy
may be experimental determined for non-relativistic process, i.e. process
which involve  the kaon system.} 

In a recent paper~\cite{OK}, it was obeserved that the equal velocity 
prescription for neutrino oscillations is forbidden because for known 
production processes 
$E_{\1} / E_{\2} \approx 1$ while $ M_{\1} / M_{\2}$ may be extremely 
small or extremely large. Nevertheless, almost equal neutrino 
masses are {\em not} ruled out experimentally. In fact,  
for the solution of
the solar neutrino problem through the matter-enhanced neutrino oscillations,
we have $\Delta M_{\so}^2 \approx 10^{\mi \5} \mbox{eV}^2$ [ for the vacum 
oscillations $\Delta M_{\so}^2 \approx 10^{\mi \1 \0} 
\mbox{eV}^2$]~\cite{EX1}.  
The explanation of atmospheric neutrino experiments through the neutrino 
oscillations requires $\Delta M_{\at}^2 \approx 10^{\mi \3} \mbox{eV}^2$
~\cite{EX2}. The hypothesis $\Delta M \ll M_{\m}$, true for the kaon 
system, implies $M_{\m} \gg 2.2 \times  10^{\mi \3}  \mbox{eV}~[ \, 
0.7 \times  10^{\mi \5}  \mbox{eV} \, ]$ for solar neutrinos and
$M_{\m} \gg 2.2 \times  10^{\mi \2}  \mbox{eV}$ for atmospheric neutrinos.
If $\Delta M \ll M_{\m}$, we get $M_{\1} / M_{\2} \approx 1$
which in the equal velocity prescription implies $E_{\1} / E_{\2} \approx 1$.
Thus, to recover the standard formula for mass oscillations, we 
need almost equal 
masses in addition to the {\em exacxt} common velocity scenario. In all
the other cases, we find a doubling of the oscillation phase.

We also observe that in the scenario of different velocities, by assuming 
$L_{\1}=L_{\2}=L$, $t_{\1}$ and $t_{\2}$ could be {\em significantly}
different, the extreme case is seen in~\cite{AHL1,AHL2}, then there is no
interference. The predictions of the equal velocity scenario is therefore
dramatically different from the other scenarios in long baseline experiments.

In our discussion, we have used plane wave amplitudes as 
approximations for our 
calculations. A complete
understanding  of neutrino oscillations requires the treatment 
of localization of the microscopic process by which a neutrino is produced 
and detected. This localization, appropriately described by a 
wave packet treatment, is essential to give a 
significance to the distance from source to measuring apparatus. 
In such a picture, the flavour eigenstate is created  not as a simple
two state system but rather as a superposition of two wave packets, one for 
each mass eigenstate. The coherence properties of 
the neutrino flux have been examined in terms of the {\em length of
the wave packet} resulting from the electron capture process, first by
Nussinov~\cite{NUS} and more recently re-examined  by him and 
collaborators~\cite{KIE}. In neutron physics, where the coherence properties 
of particle beams can be particularly 
well studied, there have been several discussions as to whether and how it 
could be possible to determine or observe the wave packet properties of a 
beam~\cite{BG}. In recent papers~\cite{KIE,STO} we also find an interesting 
discussion about the impossibility of telling the difference between beams
with  the same energy spectrum  consisting of a mixture of long and short 
wave packets. Unfortunately, {\em it is not clear what determines the 
size of the wave packet at the moment of creation or even if it makes 
sense to talk of a precise time of creation}~\cite{RIC}.
So, a clear and consistent discussion of neutrino oscillations by wave packets
still represents an open question in this research field~\cite{TSAI,TSAI2}.

After the completion of this work, two of the authors 
have reconsidered the known data upon the neutral kaon system~\cite{SdL}. 
The initial 
objective of this revision was to locate in the early papers~\cite{KS1,KS2} 
upon this 
subject and the origin and justification of the so-called standard oscillation 
formula. 
	
As shown in equation (\ref{soc}), the standard formula is obtained if 
$E_{\1} \approx E_{\2} \approx E_{\m}$, 
that is, in the case of   $\Delta M \ll M$ 
(valid for the neutral kaons but improbable for neutrinos) and furthermore if 
one assumes the {\em equal velocities} hypothesis. This equal velocities 
hypothesis indeed dominates the literature including the most recent 
papers used by the 
Particle Data Group~\cite{DP} for its estimate of  $\Delta M$ even if these 
are not based upon oscillation measurements. This fact is deducible from the 
appearance of a unique proper time, $\tau$, multiplying the 
difference in mass in the appropriate formulas, such as for 
asymmetries in charged Kaon semi-leptonic decays.
We anticipate some of the findings of this research. 

In 
particular a long-standing ``paradox'' in the literature may have as a 
solution the factor two discussed here.  
In the paper by Fujii et. al.~\cite{FUJ} upon neutral kaon decays, 
we read ``... there is a marked tendency for experiments of the first 
type [oscillations] to give significantly higher values [for $\Delta M$] than 
those of the second type [regeneration] ...". 
	
We begin by noting that these higher values, 
within the experimental errors, are larger by about a factor of two. 
We also recall that the regeneration experiments are analyzed with the 
hypothesis of equal energies for the outgoing kaon mass eigenstates. 
Indeed, this fact can be found in the book by Okun~\cite{OKUN}. 
Our observation is simply that if 
the original neutral kaon is not a common velocity scenario, then an 
overestimate of $\Delta M$  by a factor of two follows automatically from 
the incorrect use of the standard oscillation formula. 
Indeed the fitted $\Delta M$ would then have to 
compensate the extra factor of two in the denominator. 
This fact was first noted by Srivastava et al.~\cite{SRI2}.

This example shows that in the neutral kaon system there is already 
evidence for the questions posed in this paper. 
Neutral kaons also offer the practical possibility of performing experiments 
with non-relativistic particles and hence of distinguishing between the 
various scenarios discussed in this work. Already we can suggest that an 
interesting experiment would be the $\Delta M$ derived with regenerated 
kaons in two alternative studies one with time-space oscillations of the 
kind considered here and the other by forward intensity measurements or 
interference effects due to regeneration in two or more plates. 
In this way one is certain that the scenario, even if in doubt, is the 
same for the two experiments.

\acknowledgement{
The authors acknowledge useful discussions with the members of the MACRO 
group of Lecce University and the GEFAN group of S\~ao Paulo-Campinas 
collaboration and in particular, they wish to thank 
G.~Mancarella and M.~Guzzo for the critical reading of this 
manuscript and for their very useful comments. The authors are grateful
to Prof.~D.~V.~Ahluwalia for drawing their attention to ref.~\cite{OK}.
}

\end{document}